\documentclass[aps,prl,twocolumn,showpacs,nofootinbib]{revtex4}
\usepackage{graphicx}
\usepackage{amssymb}
\usepackage{epstopdf}
\usepackage{subfig}

\usepackage{dcolumn}
\usepackage{bm}
\usepackage{epsfig}

\def\ba{\begin{array}}
\def\ea{\end{array}}
\def\be{\begin{equation}}
\def\ee{\end{equation}}
\def\ben{\begin{equation} \nonumber}
\def\een{\end{equation}}
\def\baray{\begin{eqnarray*}}
\def\earay{\end{eqnarray*}}

\def\up{\uparrow}
\def\down{\downarrow}

\def\gradvec{\vec{\nabla}}
 
\def\xvec{{\bf x}}

\def\pvec{{\bf p}}
\def\phivec{\vec{\phi}}
\def\ghat{\hat g}

\begin{document}

\title{On the New Model of Non-Fermi Liquid for High Temperature Superconductivity}
\author{S.-H. Henry Tye}
\affiliation{Newman Laboratory for Elementary Particle Physics,
Cornell University, Ithaca, NY 14853, USA. 
}

\date{ May 27, 2008}

\begin{abstract}
The importance of introducing the SO(5) singlet into the  new novel model of non-Fermi liquid proposed by LeClair and collaborators is pointed out.
This SO(5) singlet order parameter leads to a phase transition (or cross-over) between the normal metal and the pseudogap region in the  phase diagram for the anti-ferromagnetic (AF) phase and the (d-wave) superconducting (SC) phase. A non-zero value for this order can increase substantially the critical temperature of the SC transition. 
\end{abstract}


\maketitle


It is well-known that the Landau Fermi liquid model does not describe the properties of the high $T_c$ superconductivity \cite{Anderson}. In search of an alternative model, the concept of the renromalization group (RG) flow and Wilson low energy effective field theory should prove useful \cite{Tasi}. Let $d$ be the number of spatial dimensions of the system we are interested in. In the action formulation, in units where $\hbar$, $k_B$ and the limiting speed are set to unity,
all terms in the Langrangian density has dimension $D=d+1$.
A contact interaction term (operator) with (mass) dimension $n<D$ (so the coupling has dimension $D-n$) is relevant since it can dominate in the low energy limit. Such interactions are relevant operators. Interaction
operators with dimensions higher than $D$ are irrelevant in low energy and so can be ignored in leading approximations. Interaction operators with dimensions equal to $D$ are marginal. Their low energy relevance depends on whether they are asymptotically free (becoming strong at low energy scales) or not. 

If we treat the quasi-particle (electron) field $\chi$ with respect to the Fermi surface as a Dirac-like 
spin $1/2$ fermion, the kinetic term of $\chi$ involves only one derivative, so $\chi$ has classical 
dimension of $[\chi] = d/2$.
The 4-Fermi interaction has naive classical dimension $2d$. In $d=3$, the effective dimension of the Cooper pair interaction (including the $\delta$-function for momentum conservation) is reduced from 6 to 4, so the BCS interaction is marginal \cite{Tasi}. As is well known, it plays a crucial role in low $T_c$ (i.e., normal) superconductivity. In the $d=2$ case for cuprates as high $T_c$ superconductors, data clearly shows that the BCS theory does not apply. The classical $[\chi] = d/2=1$, so the naive 4-Fermi interaction in the Landau Fermi liquid model has dimension $4$, which makes it irrelevant. Apparently we are left with no known relevant or marginal interaction to include. Such a theory leads to free electrons, i.e., normal metal.

\begin{figure}[t] 
\begin{center}
\hspace{-10mm}
\includegraphics[width=8cm]{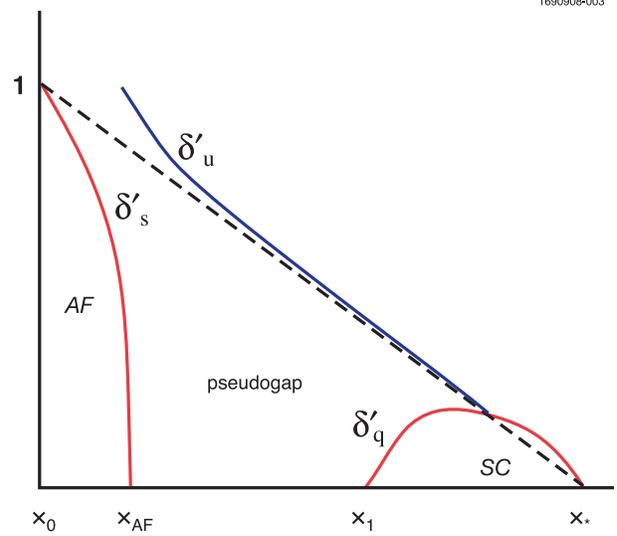} 
\end{center}
\caption{The schematic phase diagram as a function of $x=1/\ghat$, which is proportional to the hole doping, $h(x) \propto  x - x_0$. $x_*=1/\ghat_* \simeq 8$ is the renormalization 
group fixed point.
The y-axis also measures the critical temperatures.
The gap $\delta_u'$ (solid dark blue curve) is the pseudogap  (and/or the spin gap) transition discussed in the paper. The dashed line indicates the scaling $r=\Lambda/\Lambda_c$.
The presence of $\delta_u'$ enlarges the 
gap $\delta_s'$ for the anti-ferromagnetic (AF) phase and 
the gap $\delta_q'$ for the $d$-wave superconducting (SC) phase (solid red curves) with respect to that obtained in Ref.\cite{LeClair3}.
} 
\end{figure} 

In the novel work by LeClair and collaborators \cite{LeClair1,LeClair2,LeClair3}, 
an alternative $\chi$ kinetic term with 2 derivatives (of Klein-Gordon type) is proposed and analyzed.
In this model, the $[\chi] = (D-2)/2=1/2$ and the 4-Fermi interaction (with classical dimension 2) is relevant. Its strong interaction is crucial in deriving the phase diagram. If the model captures the key features of high $T_c$ superconductivity, one can incorporate more structures into it. Here our discussion closely follows Ref.\cite{LeClair3}, which 
shows 2 phase transitions : the anti-ferromagnetic (AF) and the ($d$-wave) superconducting (SC). In this paper, we show that there is a natural (automatic) extension of the basic concept to include another order parameter, which leads to a third phase transition. This is identified with the pseudogap transition. In contrast to the AF and SC phases, which have spin and charge (respectively) properties, the pseudogap phase does not have any distinguishing feature 
and may be considered as a cross-over. This completes the construction of the basic model.

This novel model has a SO(5) symmetry \cite{Zhang}, and the order parameters in the vector {\bf 5} representation has been studied in Ref.\cite{LeClair3}. The neutral order parameters in {\bf 5} lead to the anti-ferromagnetic (AF) phase and the charged order parameter in {\bf 5} leads to the superconducting (SC) phase.  In this paper, we include the SO(5) singlet order parameter. 
We argue that it is responsible for the pseudogap. Its condensate also raises the other critical temperatures, in particular that for the superconducting transition. In principle, the overall features of the phase diagram are essentially parameter-free. (In practice, some parameters remain to be determined and others to be better calculated.) Quantitative properties requires the value of the dimensionless coupling $\ghat$ at the energy-momentum cutoff scale $\Lambda_c$, $\ghat_0=\ghat (\Lambda_c)$. It is natural to take 
$\Lambda_c$ to be the inverse of the lattice spacing $a$. To obtain the critical temperatures,  one also needs to know the Fermi velocity $v_F$. 

Around the Fermi surface at less than half-filling, the model is given by
\baray
\label{L1}
S = \int  dt \, d^2 x ~ ( \sum_{\alpha} \partial_t \chi^-_{\alpha} \partial_t \chi^+_{\alpha} - v_F^2 
\gradvec \chi^-_{\alpha} \cdot \gradvec \chi^+_{\alpha} ) \\
-  8 \pi^2 g  ~ \chi^-_\up \chi^+_\up \chi^-_\down \chi^+_\down
\earay
where $\alpha= \up, \down$ and the $\pm$ indices on the fields
$\chi^\pm$ correspond to electric charge. The mode expansions of
the fields are ($\omega_\pvec = \sqrt{\pvec^2}$ and 
$p\cdot x \equiv \omega_\pvec t - \pvec \cdot \xvec $),
\baray
\nonumber
\chi^-_{\alpha} (\xvec ,t )  &=& \int \frac{d^2 \pvec}{2 \pi \sqrt{2\omega_\pvec}}
(  a^\dagger_{\alpha \pvec}  \, e^{-i p \cdot x }   +  b_{\alpha \pvec} \,  e^{i p \cdot x}  ) 
\\
\label{Mode}
\chi^+_{\alpha} (\xvec ,t )  &=& \int \frac{d^2 \pvec}{2\pi \sqrt{2\omega_\pvec}}
( - b^\dagger_{\alpha \pvec}  \, e^{-i p \cdot x }   +  a_{\alpha \pvec} \,  e^{i p \cdot x}  ) 
\earay
With $ \{\chi^{\mp}_{\alpha} , \dot \chi^{\pm}_{\beta} \}= 
\pm i \delta_{\alpha \beta}$ and $\{\chi_i , \chi_j\}=0$, the $(a, b, a^\dagger, b^\dagger)$ are the normal electron and hole annihilation and creation operators. The Hamiltonian $H$ is pseudo-hermitian ($H^\dagger = CHC$ where $C$ is a special unitary operator satisfying $C^\dagger C= C^2=1$ and distinguishes particles and holes: $CaC=a$, $CbC=-b$ and $(\chi^-)^{\dagger c}=C(\chi^-)^\dagger C=\chi^+$), ensuring a unitary time evolution and real eigenvalues \cite{Pauli,LeClair2}. 
Here the relevant 4-Fermi contact interaction is unique, up to the strength of the coupling $g \ge 0$, which has dimension $1$. The coupling $g(\Lambda)$ is scale dependent, so it is convenient to introduce the dimensionless coupling $\ghat (\Lambda) =  g(\Lambda)/\Lambda=1/x$. Ref.\cite{LeClair2} finds
\be
\label{beta}
\beta(\ghat) = \frac{d\ghat}{d\ln \Lambda} = -\ghat + 8 \ghat^2 + . . .
\ee 
which has an attractive fixed point at $\ghat_*=1/x_* \simeq1/8$. Suppose $g(\Lambda_c)=\ghat_0=1/x_0$. 
Assuming that $\beta(\ghat)$ has no other fixed points between $\ghat_0 > \ghat >\ghat_*$, we have
\be
\label{runningg}
\ghat (\Lambda)  \simeq \frac{\Lambda_c \ghat_0}{\Lambda +8(\Lambda_c -\Lambda)\ghat_0}
\ee
Hole doping goes as $h \propto x - x_0$, so we are interested in the case where $\ghat_0 \gg \ghat_*=1/8$ and study the behavior of the theory as it flows towards $\ghat_*$ as $\Lambda \rightarrow 0$.

The model has an automatic SO(5) symmetry \cite{Zhang}. 
Treating the $\chi$s as the spinor representation {\bf 4} of SO(5), 
we see that {\bf 4} x {\bf 4} x {\bf 4} x {\bf 4} = {\bf 1} + ...
where {\bf 1} is a SO(5) singlet. The interaction operator is such a singlet.
Now we can write
{\bf 4} x {\bf 4} = {\bf 1}  + {\bf 5} + {\bf 10},
where {\bf 1}, {\bf 5}  and {\bf 10} are the singlet, the vector and the adjoint representation of SO(5), respectively. That is, one can write the bilinears of $\chi$ fields in terms of pair fields or order parameters. 
The singlet {\bf 1} is given by
\be
\phi_0 = \frac{1}{\sqrt{2}}(\chi^-_\up \chi^+_\up + \chi^-_\down \chi^+_\down)
\ee 
and the vector {\bf 5} is given by
\baray
\vec{\Phi} &=& (\phivec,  \phi^+_e , \phi^-_e)
=(\phi_+ , \phi_- , \phi_z , \phi^+_e , \phi^-_e) \\\nonumber
&=& (\chi^-_\up \chi^+_\down, \chi^-_\down \chi^+_\up, \frac{1}{\sqrt{2}}(\chi^-_\up \chi^+_\up - \chi^-_\down \chi^+_\down), \chi^+_\up \chi^+_\down, \chi^-_\down \chi^-_\up )
\earay
where $\phi_\pm = (\phi_x \pm i\phi_y)/\sqrt{2}$. 
In terms of the subgroup $SU(2) \times U(1)$ of SO(5), 
the $\phivec = \chi^- {\vec{\sigma}} \chi^+ /\sqrt{2}$  
is an electrically neutral $SU(2)$ vector 
and $\phi^\pm_e$ are Cooper pair fields of charge $\pm 2$ 
which are $SU(2)$ spin singlets. Note that
\be
\label{Phi2}
\vec{\Phi} \cdot \vec{\Phi} = 2\phi_+ \phi_- + \phi_z^2 - 2\phi^+_e \phi^-_e
\ee 
where the "$-$" sign follows from pseudo-hermiticity.
Due to the Grassmanian property of the $\chi$s, the interaction Hamiltonian
can be written in terms of any combination of the order parameters,
\baray
H_{\rm int} &=&  \int d^2x  ~ 8 \pi^2 g  ~ \chi^-_\up \chi^+_\up \chi^-_\down \chi^+_\down
= -\int d^2x \frac{8 \pi^2 g}{5} ~ \vec{\Phi} \cdot \vec{\Phi} \\\nonumber
&=& - \int d^2x ~8 \pi^2 g \phi_z^2 = + \int d^2x ~8 \pi^2 g \phi_0^2 = ...
\earay
However, the interaction Hamiltonian $H_{\rm int}$ can now be generalized in a manifestly $SO(5)$ invariant fashion to allow more structure to the model :
\be
\label{so5int}
\int d^2x d^2x' ~\left[ \vec{\Phi}(x) \cdot \vec{\Phi} (x') V_5 (x, x')
 + \phi_0(x) \phi_0(x') V_0 (x,x') \right]
 \ee
where a marginal $\phi_0 \vec{\Phi} \cdot \vec{\Phi}$ term is also possible.
Electromagnetic couplings of $\phi_e^{\pm}$ imply that the SO(5) symmetry is valid only approximately. 
Since the $\chi$s are Grassmanian, a {\bf 10} must be constructed with a derivative, e.g., $Q \sim \chi \dot \chi$ (they form the SO(5) generators); but its dimension $[Q]=2$, so a singlet interaction of the form $\sum Q_{ij} Q_{ij}$ is irrelevant under naive power counting. 
(An interaction $\sim$ {\bf 5} x {\bf 5'} where {\bf 5'} contains a derivative will be marginal, but such a singlet breaks Lorentz properties.) So the model (\ref{so5int}) is the most general relevant interaction consistent with the SO(5) symmetry (based on naive power counting). However, it is important to note that, under appropriate conditions (as in some momentum subspaces), higher dimensional operators may become relevant/marginal and so must be included (e.g., Cooper pairs in the BCS model and the d-wave here). Some of these higher dimensional SO(5) symmetric operators (e.g., the d-wave here) would be generated automatically via quantum corrections.

Introducing the $\phi_0^2$ operator is the new ingredient beyond the analysis of 
Ref.\cite{LeClair3}, so we shall focus on the dynamics of this term. 
Although classically, both $[\phi_0]=1$ and $[\vec{\phi}]=1$, their anomalous dimensions have been evaluated in Ref.\cite{LeClair2}. At the RG fixed point, $[\phi_0]\simeq 5/8$ and $[\vec{\phi}]\simeq3/2$ in the $\epsilon$-expansion. This suggests that quantum effects will make the $\phi_0^2$ interaction more relevant, while pushing the $\vec{\Phi} \cdot \vec{\Phi}$ towards marginality. In this sense, the $\phi_0^2$ interaction is the most relevant and important one.

Following the standard Hubbard-Stratonovich approach, we introduce the auxiliary fields $\vec{s}$, $q^{\pm}$ and $u$ coupled to the order parameters with the action
\baray
S_{\rm aux} &=& \int dt d^2x ~ +\sqrt{2} u \phi_0 +\sqrt{2} \vec{s}\cdot \vec{\phi}+ q^+\phi_e^- 
+  q^-\phi_e^+  \\\nonumber
 &&- \frac{1}{8 \pi^2 g_u}u^2   - \frac{1}{8 \pi^2 g_s} \vec{s}\cdot \vec{s} - \frac{1}{8 \pi^2 g_q} q^+ q^-  
\earay
Variations $\delta S_{\rm aux} =0$ imply
\baray
 u= 8 \pi^2g_u \phi_0/\sqrt{2},  \quad \vec{s}=8 \pi^2g_s \vec{\phi}/\sqrt{2}, \quad q^{\pm}= 8\pi^2 g_q \phi_e^{\pm},
\earay
Substituting these back into the action reproduces the original model if
\be
\label{couplings}
- g_u/2 + 3g_s/2  -g_q =g
\ee
This relation is valid only classically. SO(5) symmetry implies $g_s=-g_q$. 

Instead, integrating out the $\chi$ fields (in euclidean space) yields an effective potential
in terms of $\vec{s}$, $q^{\pm}$ and $u$,
\baray
V_{\rm eff} &=& \frac{1}{8 \pi^2 g_u} u^2 + \frac{1}{8 \pi^2 g_s} \vec{s} \cdot \vec{s} +
 \frac{1}{8 \pi^2 g_q} q^+ q^- \\\nonumber
&& - \int \frac{d^3p}{(2 \pi)^3} \ln \left( (p^2 +m^2 -u)^2 + q^+q^- -  \vec{s} \cdot \vec{s} \right)
\earay
where the mass $m$ should be interpreted as the temperature $m \propto T$. 

Varying $V_{\rm eff}$ with respect to $u$, $\vec{s}$ and $q^{\pm}$ yields the gap equations respectively,
\be 
\label{gapequ}
u  =  - 8 \pi^2 g_u   \int  \frac{d^3p}{(2\pi)^{3}}
\, \frac{p^2 + m^2 -u}{(p^2 + m^2-u)^2 +q^+q^- -\vec{s}^2} 
\ee
\be
\label{gapeqs}
\vec{s}  =  - 8 \pi^2 g_s   \int  \frac{d^3p}{(2\pi)^{3}}
\, \frac{\vec{s}}{ (p^2 + m^2-u)^2 +q^+q^- -\vec{s}^2} 
\ee
\be
\label{gapeqq}
q^{\pm}  =  8 \pi^2 g_q   \int  \frac{d^3p}{(2\pi)^{3}}
\, \frac{q^{\pm}}{ (p^2 + m^2-u)^2 +q^+q^- -\vec{s}^2} 
\ee
Since $u$, $s=|\vec{s}|$ and $q=q^{\pm}$ have dimension 2, we introduce the dimensionless $\delta_j$ :
$u = \delta_u^2 \Lambda_c^2$, $s=\delta_s^2 \Lambda_c^2$ and $q=\delta_q^2 \Lambda_c^2$. 
Here, the $\delta_j$ are energy scale ($\Lambda$)-dependent.
The 3 gap equations are coupled to one another, so a general analysis is a little complicated.
The $\vec{s}$ gap equation and the $q^{\pm}$ gap equations have been derived and analyzed in Ref.\cite{LeClair3}.
So let us consider the $u$ gap equation (\ref{gapequ}) here. 

Since quantum effects render the $\phi_0^2$ interaction operator more relevant than the others, it is reasonable to first treat the $u$ gap equation ignoring possible $s$ and $q$ gaps. Then one can study the $s$ and $q$ gap equations treating the solved $u$ as a background.
Let $m=s=q=0$, so Eq.(\ref{gapequ}) simplifies to
\be
\label{usoln}
u= -4 g_u \int^{\Lambda_c}_0 dp \frac{p^2}{p^2-u} 
\ee
Here $u \rightarrow 0$ as $g_u \rightarrow 0$; so $u \rightarrow 0$ is the solution as $g_u  \rightarrow 0$. Note that this equation (\ref{usoln}) is different from the other 2 gap equations in that RHS of Eq.(\ref{usoln}) is not proportional to $u$. For $g_u >0$,
the $s$-wave attractive interaction yields a gap at scale $\Lambda$ for $\delta_u>1$, 
\be
\delta_u = \frac{4g_u}{\Lambda_c} \left(\tanh^{-1} 1/\delta_u  -\frac{1}{\delta_u} \right)
\ee
For large $g_u/\Lambda_c$, we have $\delta_u \simeq ({4 g_u}/{3 \Lambda_c})^{1/4}$.
For $\delta_u \gtrsim1$, we have
$\delta_u = 1 + 2e^{-\Lambda_c/2g_u} \rightarrow 1$
if $g_u$ decreases as $\Lambda$ decreases, as expected. 
Including classical scaling, we have $\delta'_u \sim   \delta_u r =   \delta_u \Lambda/\Lambda_c$. This is shown in 
Fig 1.

However, $g_u$ may be negative. Following Eq.(\ref{couplings}), one may expect $g_u \sim - g$. In this case, we may solve Eq.(\ref{usoln}) for $u$ perturbatively. To leading order in $g$, we now have
$u \sim 4g\Lambda_c =  4{\hat g} \Lambda \Lambda_c$
In this crude approximation, we have
$\delta_{u}'  = \sqrt{u/\Lambda_c^2}  \sim   \sqrt{4 {\hat g} \Lambda/{\Lambda_c}}$, where $\hat g$ is approximately given in Eq.(\ref{runningg}). Because of the form of the equation (\ref{usoln}), this is more like a cross-over than a phase transition. Note that irrespective of the sign of $g_u$, we have a positive non-zero $u$ solution.

The presence of a non-zero $u$ does not break the SO(5) symmetry, so no Goldstone mode is introduced. Since it does not break the $SU(2)$ spin or the electromagnetic U(1) symmetries, its detection is more subtle than the other 2 phase transitions. It should show up in the properties of, e.g., the specific heat and the paramagnetic susceptibility \cite{pseudo}.

The $\delta'_j$ (or $\delta_j$) ($j=u,s,q$) gaps are scale-dependent. Analogous to the anomalous mass dimension $\gamma_m$ for the mass, we expect an anomalous gap dimension, $\gamma_j = \frac{\partial \ln \delta_j}{\partial \ln \Lambda}$,
whose 2-loop result
\cite{LeClair2} may not be that informative for strong couplings.
Clearly their scaling 
behaviors remains to be determined. So far, except for the running of the coupling, 
classical scaling ( as a function of $r= \Lambda/\Lambda_c$ has been used. 

Next, we are interested in the determination of the pseudogap temperature $T_P$, which is related via the $u$ gap equation (\ref{gapequ}) by
$m^2-u=0$, where the mass $m= \alpha T$.  \cite{LeClair3} estimates that $\alpha \simeq 1.7$. This leads to the critical temperature for the $\delta_j'$ gap,
\be
\label{tempj}
T_j \simeq c_j\delta_j' \frac{\Lambda_c}{\alpha}  \quad \quad j=u,s,q
\ee
where $c_j$ is some normalization factor.
In Fig. 1, we see that all of the AF and SC regions are probably inside the region with a non-zero $u$ value, so we expect those regions will be modified by the presence of $u$. 

The $s$ gap equation (\ref{gapeqs}) is attractive for weak doping, so it has a solution, with a non-zero $\delta_s'$ that breaks the $SU(2)$ symmetry \cite{LeClair3}. It is shown in Fig. 1.

With $g_q$ positive, the $s$-wave $q^{\pm}$ gap equation (\ref{gapeqq}) has a repulsive potential and so does not have a solution. Physically, it is not surprising since the pairing of 2 electrons has a repulsive Coulomb force between them, which is absent in the $u$ and $s$ pairings. This follows from 
Eq.(\ref{couplings}), which comes from 
the "$-$" sign in Eq.(\ref{Phi2}), a consequence of the definition of $\phi^{\pm}_e$, which in turn is dictated by pseudo-hermiticity.
Introducing  non-constant auxiliary pair fields and one-loop contribution, Kapit and LeClair derived a momentum-dependent gap equation, which has a $d$-wave attractive channel for Cooper pairs \cite{LeClair3}.
This yields the superconducting phase, i.e., the SC region in the phase diagram
schematically shown in Fig. 1. The effective $d$-wave coupling 
$g_2 = 4\ghat^2/25 \Lambda^3$ decreases rapidly as doping is decreased (i.e., $\Lambda$ increases). Once $g_2$ becomes too small, the potential is too weak to bind, so the SC region terminates. This is the picture in the absence of a $u$ value. Now we like to make a crude estimate of the impact of a non-zero $u$ on the SC region. 

The high doping end of the SC dome region is terminated at the critical fixed point $x_*=8$. The upper SC transition at doping $h (x_*=8) = {3}/{2\pi^2}\simeq 0.15$ is second-order and corresponds to the fixed point of the renormalization group. This is universal in this model. So we expect the main effect of a positive $u$  is to raise the height of the SC dome, that is, raising the critical temperature $T_{SC}$ for the superconducting transition. Since the lower doping end is due to the weakness of the $d$-wave coupling,  the low doping end ($x_1$) may decrease some, but we do not expect it to change much. 

The (maximum) critical temperature $T_0=T_{SC}(u=0)$ in the absence of $u$ is estimated in Ref.\cite{LeClair3}, in which the reader can find the details. Restoring in Eq.(\ref{tempj}) $\hbar$ and $k_B$ and the Fermi velocity $v_F \simeq 210 ~ km/s$ \cite{vFermi} as the limiting speed, and putting in values obtained in Ref.\cite{LeClair3}, $c_q \sim 0.5$, $\alpha \sim 1.7$, $\delta_{q0}'=\delta_q'(u=0) \simeq 0.11$ and $\Lambda_c =1/a$, where the lattice spacing is $a \simeq 3.8 \times 10^{-10} m$, one obtains 
\be
 T_{SC} (u=0) = \frac{c_q}{\alpha} \frac{v_F \hbar \delta_{q0}' \Lambda_c}{k_B}= Jc_q \delta_{q0}'\sim 140^oK
\ee
where the temperature $J ={v_F \hbar}/{a\alpha k_B} \simeq 2500^oK$. The actual Neel temperature is lower than $J$, presumably due to interlayer couplings.

Looking at the effective mass term in the denominator of the $q$ gap equation (\ref{gapeqq}),
we see that $m^2 \rightarrow m^2 -u$. This means that, in the presence of a non-zero $\delta_u'$, the $q$ gap equation can go to a higher $T$ without losing its gap solution. That is, the critical temperature $T_{SC}$ is enhanced. The effect is largest for largest $\Lambda$ (smallest doping) where $g_2$ has not yet cut off the $q$ gap solution. Taking $\ghat_0 \rightarrow \infty$ and $\delta_u' \simeq \sqrt{r} \sim 0.2$,
\be
T_{SC} = J \left[(c_q\delta_{q0}')^2 + (c_u\delta_u')^2 \right]^{1/2} \simeq 280^oK
\ee
where we simply take $c_u \sim c_q$. Although this value of $T_{SC}$ is for illustrative purpose only, we 
see that a $u>0$ value (irrespective of the sign of $g_u$) enhances the $T_{SC}$.
At this doping value, the pseudogap transition temperature $T_P= Jc_u \delta_u' \simeq 250^oK$ is bigger than $T_0$.
Since $\delta_u'$ gap vanishes above $T_P$, we expect $T_{SC}$ to loosely track $T_P$ until it reaches the lower doping value where the critical temperature drops to zero (i.e., the $q$ gap disappears). 
As pointed out in Ref.\cite{LeClair3}, decreasing the lattice spacing or increasing the Fermi velocity increases the critical temperatures. We see that a more accurate determination of the scaling properties of the $\delta'_j$ (as well as $\alpha$ and $c_j$) will provide a crucial test of the model.
 
Similarly, the combination of  $q^+q^--\vec{s}^2$ in the denominators inside the gap equations (\ref{gapeqs}, \ref{gapeqq}) suggests that  binding in one channel strengthens the binding in another channel (this may happen if $\ghat(\Lambda_c) < \ghat_*$, or in the electron doped region). 
Instead of competing,  the order parameters tend to strengthen each other. In summary, the inclusion of the SO(5) invariant order completes the construction of the basic model. The interplay of the 3 order parameters yields a non-trivial phase diagram.

I thank Andre LeClair for explaining his work to me and many useful discussions. 
This work is supported by the National Science Foundation (NSF) under grant PHY-0355005.

\end{document}